\documentclass[aps,prl,twocolumn,superscriptaddress]{revtex4-2}
\usepackage{epsfig,amssymb,amsmath,amsthm,amsfonts,amsbsy,mathrsfs}
\usepackage{graphicx}
\usepackage{color}
\usepackage{bm}
\usepackage{tikz}
\usetikzlibrary{arrows,shapes,chains}

\definecolor{cream}{RGB}{222,217,201}

\usepackage[%
  colorlinks=true,
  urlcolor=blue,
  linkcolor=blue,
  citecolor=blue
]{hyperref}

\newcommand{\pj}{\phi_{\rm J}}

\begin{document}
\title{Interplay between jamming and MIPS in persistent self-propelling particles}
\author{Jing Yang}
\affiliation{Division of Physics and Applied Physics, School of Physical and
Mathematical Sciences, Nanyang Technological University, Singapore 637371}
\author{Ran Ni}
\affiliation{Chemical Engineering, School of Chemical and Biomedical Engineering,
Nanyang Technological University,  Singapore 637459}
\author{Massimo Pica Ciamarra}
\email{massimo@ntu.edu.sg}
\affiliation{Division of Physics and Applied Physics, School of Physical and
Mathematical Sciences, Nanyang Technological University, Singapore 637371}
\affiliation{CNRS@CREATE LTD, 1 Create Way, \#08-01 CREATE Tower, Singapore 138602}
\affiliation{
CNR--SPIN, Dipartimento di Scienze Fisiche,
Universit\`a di Napoli Federico II, I-80126, Napoli, Italy
}
\date{\today}

\begin{abstract}
In living and engineered systems of active particles, self-propulsion induces an unjamming transition from a solid to a fluid phase and phase separation between a gas and a liquid-like phase.
We demonstrate an interplay between these two nonequilibrium transitions in systems of persistent active particles. 
The coexistence and jamming lines in the activity-density plane meet at the jamming transition point in the limit of hard particles or zero activity.
This interplay induces an anomalous dynamic in the liquid phase and hysteresis at the active jamming transition.
\end{abstract}
\maketitle

Giant density fluctuations and collective phenomena reminiscent of equilibrium phase transitions such as flocking~\cite{vicsek1995novel}, motility induced phase separation (MIPS)~\cite{fily2012athermal,RevModPhys.85.1143,RevModPhys.88.045006,cates2015motility} and active crystallization~\cite{palacci2013living} characterize living and engineered systems of particles able to self-propel.
These phenomena emerge on increasing the strength of the self-propelling forces at the expense of other collective phenomena. 
For instance, in the prototypical hard-disk system, active forces affect equilibrium melting by first suppressing liquid/hexatic coexistence and then inducing a MIPS between a low-density gas-like phase and a higher density liquid, hexatic or crystalline phase~\cite{bialke2012crystallization, redner2013structure, digregorio2018full}. 
In three dimensions, the motility induced gas/liquid transition occurs within the gas/crystal MIPS coexistence region~\cite{omar2021phase}.



Active forces influence the glass and jamming transitions in systems that do not crystallize due to structural or energetic disorder, including cell aggregates~\cite{Henkes2011, henkes2020dense, Bi2016, garcia2015physics, Giavazzi2018, mongera2018fluid,D0SM00109K, Li2021,Lawson-Keister2021}, bacteria colonies~\cite{parry2014bacterial, Delarue2016, yang2019quenching} and polydisperse active Brownian particles \cite{Henkes2011,mandal2020extreme,mandal2020multiple,Paoluzzi2021}.
The interplay between MIPS and jamming depends on the persistence time of the active force, that influences MIPS~\cite{nie2020stability}.
In two dimensional systems of soft,  bi-disperse active Brownian disks~\cite{Henkes2011,fily2014freezing} with a `small' persistence time, jamming and MIPS appears unrelated as occurring in different regions of the density/activity plane.
Recent works~\cite{Paoluzzi2021,keta2022disordered} investigated the limit of high persistence in models differing in polydispersity, thermal noise and active velocity dynamics, reporting contrasting results. 
In~\cite{Paoluzzi2021}, MIPS and jamming (glass) stay separate in the limit of high persistence, while in ~\cite{keta2022disordered}, they approach each other.
The possible connection between MIPS and jamming thus remain elusive.

In this Letter, we demonstrate an interplay between MIPS and jamming in a two-dimensional system of active, purely repulsive particles. 
We focus on the limit of persistent particles~\cite{merrigan2020arrested,liao2018criticality,Reichhardt2014} and investigate MIPS and jamming as active forces, density, and stiffness of the particles vary.
We find that the high-density gas/liquid MIPS coexistence curve and the jamming line are separated by a small volume fraction range of liquid phase that vanishes in the limits of small activities or hard particles.
In these limits, MIPS and jamming occur together.
In the liquid phase, particle motion is correlated over the whole system for a long, size-dependent transient, during which particles do not preferentially move along the direction of their self-propelling force.
Correlations in the direction of the active forces that build up in the liquid phase induce hysteresis at the jamming transition.

We simulate a A:B $65:35$ binary mixture~\cite{bruning2008glass} of $N=16000$ particles of unit mass $m$ and diameters $d_{\rm AA}$ and $d_{\rm BB} = d_{\rm AA}/1.4$, interacting via a LJ-like n-m potential with m = n/2
\begin{equation}
V_{\rm n,m}(r) = \frac{\epsilon_{\alpha\beta}}{\rm n-m}\left[{\rm m} \left(\frac{d_{\alpha\beta}}{r}\right)^{\rm n}-{\rm n} \left(\frac{d_{\alpha\beta}}{r}\right)^{\rm m}\right]
\end{equation}
truncated in its minimum $d_{\alpha\beta}$. 
Hence, the interaction is purely repulsive.
We fix $\epsilon_{\rm AA} = 1$, $\epsilon_{\rm BB} = 0.5\epsilon_{\rm AA}$ and $\epsilon_{\rm AB}=1.5\epsilon_{\rm AA}$, and set $d_{AB} = 1/2(d_{\rm AA}+d_{\rm BB})$ and n = 12 if not otherwise stated.
The area fraction is $\phi = N L^{-2} \langle a \rangle$, with $\langle a \rangle$ the average particle area~\footnote{$\langle a \rangle = \frac{\pi}{4}[0.64 d_{\rm AA}^2+0.35 d_{\rm BB}^2])$} and $L$ the linear size of our square simulation domain.
The equation of motion for particle $i$ is
\begin{equation}
m \ddot{\mathbf{r}}_{i}= \sum_j\bm{f}_{ij}-\gamma \dot{\mathbf{r}_{i}}+\mathbf{F}_{A,i},
\label{eq:motion}
\end{equation}
where ${\bm f}_{ij}$ is the interaction force between particles $i$ and $j$, $\gamma = 1$ is a damping parameter, and $\mathbf{F}_{A,i}=F_A\mathbf{e}_i$ the self-propelling force acting on the particle.
In the range of parameters we consider, the damping parameter $\gamma$ is large enough for inertial effect to be negligible, as we will explicitly demonstrate.

The active forces have magnitude $F_A$ and fixed random orientations $\mathbf{e}_i$ which we chose with the constraint $\sum \mathbf{e}_i = 0$ to avoid the motion of the centre of mass.
We indicate with $v_A = F_A/\gamma$ and $\tau_A = d_{\rm AA}/v_A$ the typical velocity and time scale set by the active force dynamics and particle size.

\emph{Zero-activity jamming transition --} 
Our model reproduces the jamming phenomenology in the absence of active forces: energy minimal configurations acquire mechanical rigidity above a jamming area fraction that depends on their preparation protocol~\cite{Chaudhuri2010, PicaCiamarra2010}.
We investigate the jamming transition by minimizing the energy of random configurations of area fraction $\phi$ via the conjugate-gradient method~\cite{OHern2002}, using a protocol that is not affected by inertia as our active particle simulations. We estimate the jamming volume fraction (not shown) to be $\pj \simeq 0.828$.

\emph{Interplay between MPIS and jamming --} 
\begin{figure}[!t]
\centering
\includegraphics[angle=0,width=0.42\textwidth]{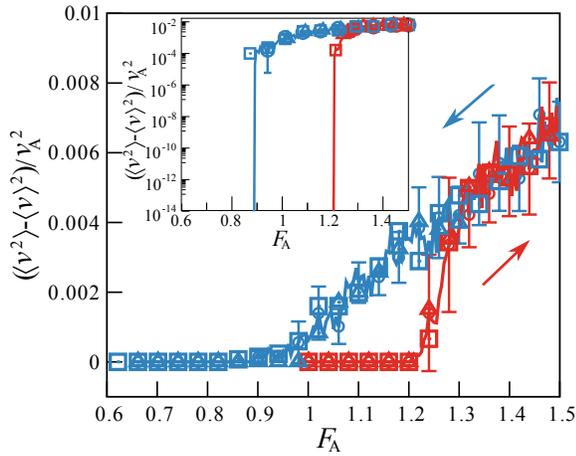}
\caption{
Dependence of the velocity fluctuations on the active force as this decreases (blue) or increases (red) at force rates $|dF_A/dt| = 10^{-6}$ (circles), $5\cdot 10^{-7}$ (triangles), $2\cdot 10^{-6}$ (squares). 
The volume fraction is $\phi = 0.98$, and the stiffness exponent is n = 12.
The abrupt variation of the kinetic energy upon jamming/unjamming illustrated in the inset allows for the unambiguous identification of the active force values of the jamming and unjamming transitions.
\label{fig1}
}
\end{figure}
In the presence of active forces, the system transitions from a flowing to a jammed regime as the density increases or the magnitude of the active forces decreases. 
We locate this transition by determining the state point $(\phi,F_A)_J$ at which the kinetic energy of the system vanishes as $F_A$ slowly decreases at a constant $\phi$, as illustrated in Fig.~\ref{fig1}.
We repeat this study four times at each considered $\phi$ to estimate the average active force at jamming.

Besides controlling the jamming transition, active forces also induce MIPS between a gas and a liquid-like phase.
Here, we cannot determine the MIPS phase boundary via the study of the equation of state as the pressure is not well defined for persistent self-propelling particles~\cite{Solon2015a}.
Henceforth, we assess phase separation by investigating the distribution of the local volume fraction $\phi_l$ obtained by coarse-graining the system on a square grid with edge length $4\sigma_{AA}$.
We summarize the result of this standard investigation~\cite{redner2013structure,speck2014effective,nie2020stability} in the supplementary material~\cite{SM}.

\begin{figure}[!!t]
\centering
\includegraphics[angle=0,width=0.47\textwidth]{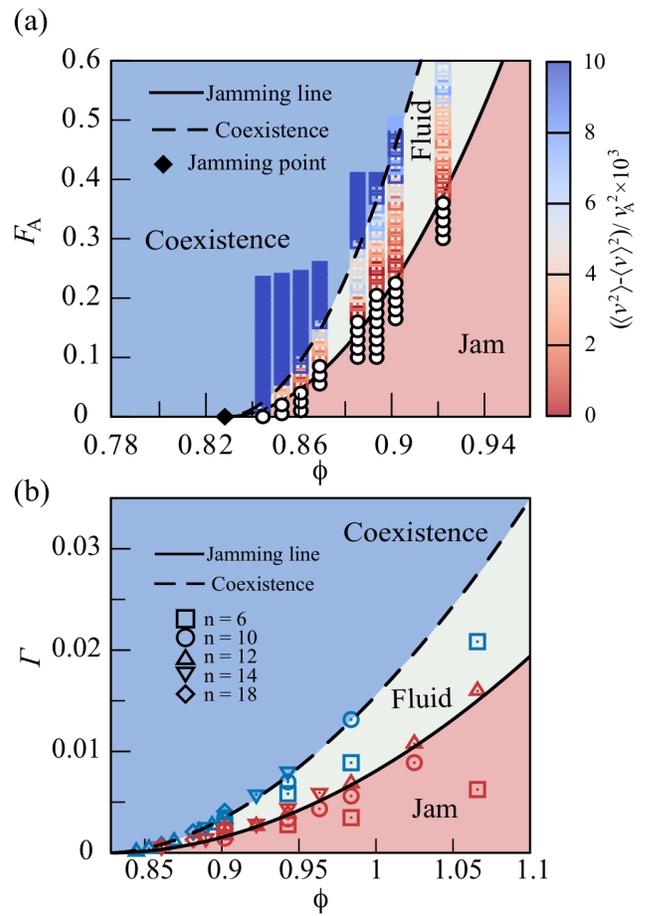}
\caption{
(a) Phase diagram for stiffness exponent n = 12.
Squares identifying phase separated (full) and homogeneous (empty) states are coloured according to the value of a scaled kinetic energy.
White circles identify jammed configurations, and the black diamond marks the jamming volume fraction, $\pj$.
The jamming transition and the high-density branch of the coexistence line scales as $(\phi-\pj)^2$.
The MIPS low-density branch (not shown) is at $\phi \simeq 0.2$ and depends weakly on $F_A$.
(b) Points on the coexistence (blue) and jamming (red) lines for potentials differing in their stiffness exponent n, in the relative compression area fraction phase diagram. 
The coexistence (dashed) and jamming (full) lines refer to n = 12 as in panel (a).
}
\label{fig:diagram}
\end{figure}

The investigation of the jamming transition and the motility induced phase separation leads to the $F_A$--$\phi$ phase diagram of Fig.~\ref{fig:diagram}(a).
The jamming line $F_J(\phi)$ separating the fluid and the jammed phase and the high-density branch of the coexistence line $F_C(\phi)\geq F_J(\phi)$ are well described by power-law functional forms vanishing at the jamming volume fraction $\pj$, $F_J = A_J(\phi-\pj)^{\beta_J}$ and $F_C = A_C (\phi-\pj)^{\beta_C}$.
We find $A_J < A_C$ and $\beta_J \simeq \beta_C \simeq 2$.

This phase diagram suggests that the coexisting phases are always of gas and a liquid type, rather than of gas and jam type~\cite{mandal2020extreme}.
Indeed, in the coexistence region, we find that the high density clusters have a finite lifetime, as we discuss in the SM~\cite{SM}.

To asses the role of particles' stiffness we determine the activity/volume fraction phase diagram for interaction potentials differing in their stiffness exponent n.
We identify the jamming density as that at which the kinetic energy drops to zero as the active force slowly decreases in magnitude, and the coexistence density as the highest density at which the local volume fraction distribution is unimodal, at the considered $F_A$ value.
To compare these potentials, we evaluate the relative particle deformation induced by the active forces, which in the harmonic approximation is $\Gamma \propto \frac{F_A d}{\epsilon {\rm n}^2}$.
In the limit of small $\Gamma$, the harmonic approximation holds, and the coexistence and jamming curve of potentials with different stiffness collapse in the $\Gamma$-$\phi$ plane, as shown in Fig.~\ref{fig:diagram}(b).
As $\Gamma$ increases, the harmonic approximation breaks down, leading to an increase of the volume fraction range where the liquid phase occurs as the potential softens (n decreases).

Our investigation demonstrates that, for persistent particles, the fluid phase vanishes at the jamming point in the $F_A \to 0$ limit, or equivalently, in the limit of hard spheres, n $\to \infty$. In these limits, MIPS and jamming meet at the jamming transition point.
To explain this result, we consider in the high-density coexistence region our system appears as a dense liquid punctuated by empty cavities (see Fig.~S1~\cite{SM}), as previously observed~\cite{Wysocki2014, Bialke2015}.
The shrinking of these cavities with the active force's magnitude drives the convergence of the high-density coexistence curve to the jamming point.

The convergence of jamming and MIPS has significant consequences for the speculated analogy~\cite{Berthier2014,liao2018criticality,Morse2021} between sheared amorphous solids and dense active matter. 
While both the increase of shear stress and activity induces an unjamming transition, the unjammed phase is homogeneous in the case of shear forces while it is phase-separated in the limit of small persistent active forces.
Furthermore, in the case of shear~\cite{ciamarra2009jamming} the jamming transition line scales linearly with the overcompression $\phi-\phi_J$, in the harmonic regime, while we find it here to scale quadratically.

\begin{figure}[!!t]
\centering
\includegraphics[angle=0,width=0.48\textwidth]{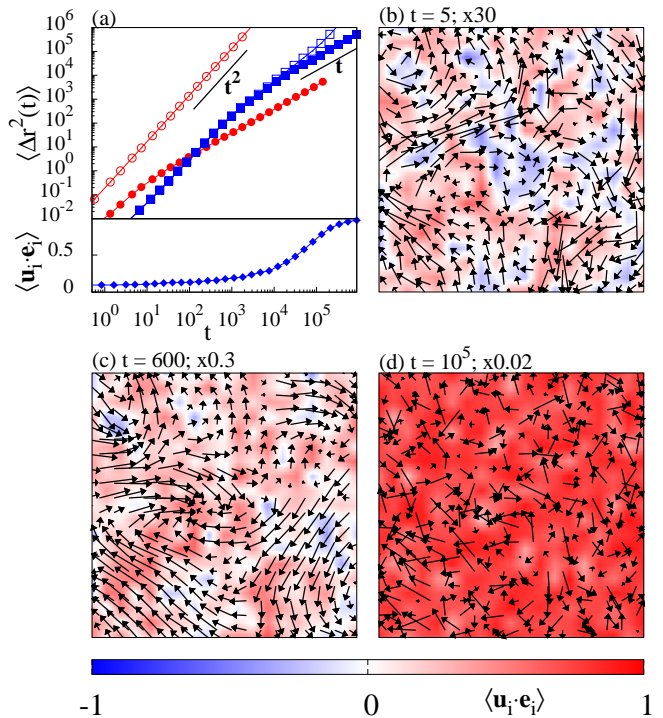}
\caption{
(a, top) Mean square displacement parallel (open symbols) and orthogonal (full symbols) to the self-propelling direction, in the gas (red, $\phi = 0.08$, $F_A = 0.5$) and the liquid (blue, $\phi = 0.9$, $F_A = 0.5$) phase, for a $N = 4000$ particle systems. (a, bottom) In the liquid phase, displacements align to the self-propelling forces after a long transient.
(b-d) Coarse grained displacement field in the liquid phase, at increasing times, scaled as indicated.
}
\label{fig:dynamics}
\end{figure}

{\it Anomalous dynamics in the liquid phase --}
The fluid phase enclosed between the coexistence and the jamming line exhibits anomalous dynamical features which we highlight by decomposing a particle' displacement in components parallel and orthogonal to its active force,
${\mathbf \Delta r}_i =  \Delta r_i {\bf u}_i = \Delta r_{i,n}{\bf e}_i + \Delta r_{i,o} {\bf  o}_i$. 
This decomposition allows us to investigate the parallel $\langle \Delta r_{i,n}^2 (t) \rangle$ and $\langle \Delta r_{i,o}^2 (t) \rangle$ orthogonal mean square displacements (MSD).
In the gas phase, the parallel MSD reflects ballistic dynamics with an effective volume fraction dependent velocity, while the orthogonal MSD reveals diffusive dynamics induced by the interparticle collisions~\cite{nie2020stability}, as we illustrate in Fig.~\ref{fig:dynamics}(a).

Surprisingly, in the high-density liquid phase parallel and orthogonal dynamics are ballistic and identical for a long transient.
During this transient, displacements are not aligned to the self-propelling forces and $\langle {\bf u}_i \cdot {\bf e}_i \rangle < 1$, as in Fig.~\ref{fig:dynamics}(a, bottom).
The orthogonal MSD transitions to a diffusive regime only after reaching order $L^2$, as we verified via a finite-size investigation (not shown).
These results originate from the transient organization of the flow pattern in large structures with a size comparable to that of the system, as in Fig.~\ref{fig:dynamics}(b)-(d).
These structures are not the turbulent-like vortices of other active matter systems~\cite{Alert2022}, as their dimension is fixed by the system size rather than by motility and interaction parameters.

\emph{Jamming vs unjamming --} 
\begin{figure}[!!t]
\centering
\includegraphics[angle=0,width=0.48\textwidth]{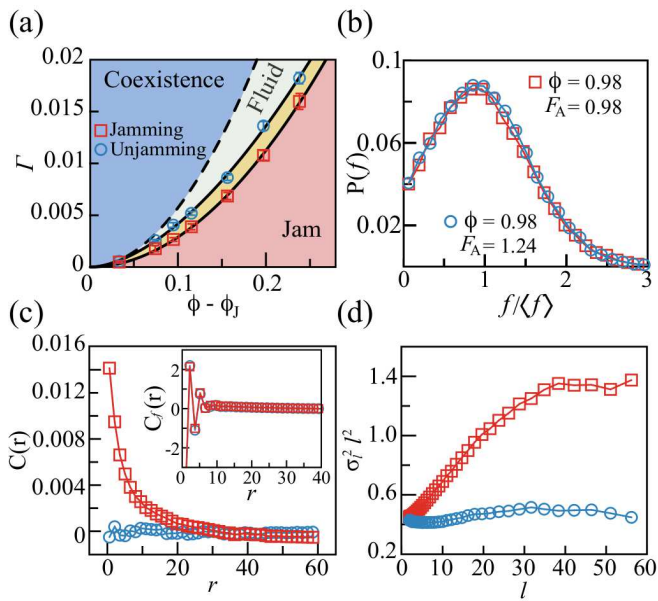}
\caption{
(a) The jamming phase diagram of Fig.~\ref{fig:diagram} ($n = 12$) with the addition of the unjamming line. 
The system could flow or be jammed in the yellow shaded region, depending on the preparation protocol.
(b) Probability distribution of the interparticle forces on representative state points on the jamming (red squares) and on the unjamming line (blue circles).
(c) correlation function of the interparticle force (inset) and of the self-propelling forces (main panel).
(d) scaled fluctuations $\sigma^2 l^2$ of the sum of the magnitude of the active forces found in square regions of linear size $l$.
\label{fig:hysteresis}
}
\end{figure}
A liquid configuration jams as the magnitude of the active forces becomes smaller than a threshold.
Similarly, a jammed configuration starts flowing if the magnitude of randomly oriented active forces overcomes an unjamming threshold.
Surprisingly, the unjamming threshold is larger than the jamming one, as apparent in Fig.~\ref{fig1}.
The difference between these two thresholds decreases with the volume fraction and vanishes at the jamming point, where the jamming and unjamming lines meet, as illustrated in Fig.~\ref{fig:hysteresis}(a).
While this distinction between jamming and unjamming resembles the inertia induced hysteresis occurring in sheared granular media~\cite{ciamarra2009jamming},
inertia in our system is negligible.
Henceforth, the observed distinction  implies differences in the configurations on the jamming and unjamming lines, which we unveil by investigating the features of their force network.

We find that configurations on the jamming and unjamming line have the same interparticle forces distribution, once the forces are scaled by their average magnitude, as we illustrate in Fig.~\ref{fig:hysteresis}(b).
Active forces do not influence the force distribution as their value on the jamming/unjamming lines is a small fraction ($\simeq 1/10$) of the typical interparticle force.
We investigate correlations in the interparticle forces by considering that each force ${\bm f}$ act at the contact point ${\bf r}_c$ of our extended interacting particles. 
We then study the correlation function between interaction forces at a distance $r=|{\bf r}_c-{\bf r}'_c|$, $C_f(r) = \langle |{\bm f}({\bf r}_c)|^q|{\bm f}({\bf r}_c')|^q \cos(2\theta) \rangle$, where $\theta$ the angle between the two forces and $q$ a parameter used to weight the contribution of forces of different magnitude to the correlation function. 
The factor of $2$ in the cosine accounts for the fact that a contact force ${\bm f}(\bf r)$ is defined up to a sign, as it could act on one of the two interacting particles.
For $q = 0$, $C_f(r)$ reduces to the two-dimensional nematic correlation function.
The inset of Fig.~\ref{fig:hysteresis}(c) reveals that $C_f(r)$ for $q = 1$ is the same on the jamming and unjamming lines.
Analogous results occur at different $q$ values, proving the absence of two-body correlations between the interaction forces.

We rationalize the difference between configurations on the jamming and unjamming lines considering that spatial correlations between the active forces build up in the liquid phase~\cite{Wysocki2014,keta2022disordered}.
If these correlations persist as the system jams, then they induce correlations in the interparticle forces as in a jammed configuration
$\mathbf{F}_{A,i} = -\sum_j {\bm f}_{ij}$.
We check this possibility by investigating the correlation function of the active force direction, $C(r)=\langle \mathbf{e}_{i}(0) \cdot \mathbf{e}_{j}(r)\rangle$ in Fig.~\ref{fig:hysteresis}(c).
At the unjamming threshold, forces are uncorrelated, so that $C(r) = 0$. 
Conversely, on the jamming line $C(r)$ only approaches zero at large distances.
As an alternative measure of correlations of the interparticle forces, we investigate the fluctuations $\sigma^2_l$ of $|\sum_{i \in l^2} \mathbf{F}_{A,i}|$, where the sum is over all particles located in square regions of linear size $l$.
At the unjamming threshold, forces are uncorrelated, and in Fig.~\ref{fig:diagram}(d) 
we find $\sigma^2_l \propto l^2$ at all $l$, as dictated by the central limit theorem.
Conversely, on the jamming line the above scaling signalling the absence of correlations only occurs for large $l$.
The results of panels (b) and (d) consistently show the existence of many-body~\cite{Zheng2021} correlations extending up to $r \simeq 30$ in the considered configuration.
We have not observed clear variations of this correlation length along the jamming line.
Henceforth, while forces on the unjamming line are uncorrelated, those on the jamming line possess many-body correlations. 

\emph{Discussion --} 
Our results demonstrate an interplay between jamming and motility induced phase separation in systems of persistent self-propelling particles.
In the $\Gamma$-$\phi$ plane, with $\Gamma$ a measure of the relative particle deformation induced by the active force, the MIPS and jamming lines meet in the $\Gamma \to 0$ limit (hard sphere or zero activity) at the jamming point.
This interplay induces surprising size effects in the dynamics of the liquid phase separating MIPS and jamming at finite $\Gamma$.
Particle motion is collective on a length comparable to that of the system for a long transient also scaling with the system size.
During this transient, particle displacements do not correlate with the directions of the self-propelling forces. 
In this liquid phase, active forces develop spatial correlations that persist as the system jam, inducing many-body correlations in the interparticle forces of jammed configurations. In the presence of a finite and large persistent time, at high density the system evolves through an intermittent avalanche dynamics~\cite{mandal2020extreme}. The huge many-body correlations in the interparticle forces we have reported may explain why these avalanches are extensive.

\begin{acknowledgments}
We acknowledge support from the Singapore Ministry of Education through the Singapore Academic Research Fund RG86/19 and RG56/21, and the National Supercomputing Centre Singapore (NSCC) for the computational resources.
\end{acknowledgments}


\begin{thebibliography}{46}%
\makeatletter
\providecommand \@ifxundefined [1]{%
 \@ifx{#1\undefined}
}%
\providecommand \@ifnum [1]{%
 \ifnum #1\expandafter \@firstoftwo
 \else \expandafter \@secondoftwo
 \fi
}%
\providecommand \@ifx [1]{%
 \ifx #1\expandafter \@firstoftwo
 \else \expandafter \@secondoftwo
 \fi
}%
\providecommand \natexlab [1]{#1}%
\providecommand \enquote  [1]{``#1''}%
\providecommand \bibnamefont  [1]{#1}%
\providecommand \bibfnamefont [1]{#1}%
\providecommand \citenamefont [1]{#1}%
\providecommand \href@noop [0]{\@secondoftwo}%
\providecommand \href [0]{\begingroup \@sanitize@url \@href}%
\providecommand \@href[1]{\@@startlink{#1}\@@href}%
\providecommand \@@href[1]{\endgroup#1\@@endlink}%
\providecommand \@sanitize@url [0]{\catcode `\\12\catcode `\$12\catcode
  `\&12\catcode `\#12\catcode `\^12\catcode `\_12\catcode `\%12\relax}%
\providecommand \@@startlink[1]{}%
\providecommand \@@endlink[0]{}%
\providecommand \url  [0]{\begingroup\@sanitize@url \@url }%
\providecommand \@url [1]{\endgroup\@href {#1}{\urlprefix }}%
\providecommand \urlprefix  [0]{URL }%
\providecommand \Eprint [0]{\href }%
\providecommand \doibase [0]{https://doi.org/}%
\providecommand \selectlanguage [0]{\@gobble}%
\providecommand \bibinfo  [0]{\@secondoftwo}%
\providecommand \bibfield  [0]{\@secondoftwo}%
\providecommand \translation [1]{[#1]}%
\providecommand \BibitemOpen [0]{}%
\providecommand \bibitemStop [0]{}%
\providecommand \bibitemNoStop [0]{.\EOS\space}%
\providecommand \EOS [0]{\spacefactor3000\relax}%
\providecommand \BibitemShut  [1]{\csname bibitem#1\endcsname}%
\let\auto@bib@innerbib\@empty
\bibitem [{\citenamefont {Vicsek}\ \emph {et~al.}(1995)\citenamefont {Vicsek},
  \citenamefont {Czir{\'o}k}, \citenamefont {Ben-Jacob}, \citenamefont
  {Cohen},\ and\ \citenamefont {Shochet}}]{vicsek1995novel}%
  \BibitemOpen
  \bibfield  {author} {\bibinfo {author} {\bibfnamefont {T.}~\bibnamefont
  {Vicsek}}, \bibinfo {author} {\bibfnamefont {A.}~\bibnamefont {Czir{\'o}k}},
  \bibinfo {author} {\bibfnamefont {E.}~\bibnamefont {Ben-Jacob}}, \bibinfo
  {author} {\bibfnamefont {I.}~\bibnamefont {Cohen}},\ and\ \bibinfo {author}
  {\bibfnamefont {O.}~\bibnamefont {Shochet}},\ }\bibfield  {title} {\bibinfo
  {title} {Novel type of phase transition in a system of self-driven
  particles},\ }\href
  {https://doi.org/https://doi.org/10.1103/PhysRevLett.75.1226} {\bibfield
  {journal} {\bibinfo  {journal} {Physical Review Letters}\ }\textbf {\bibinfo
  {volume} {75}},\ \bibinfo {pages} {1226} (\bibinfo {year}
  {1995})}\BibitemShut {NoStop}%
\bibitem [{\citenamefont {Fily}\ and\ \citenamefont
  {Marchetti}(2012)}]{fily2012athermal}%
  \BibitemOpen
  \bibfield  {author} {\bibinfo {author} {\bibfnamefont {Y.}~\bibnamefont
  {Fily}}\ and\ \bibinfo {author} {\bibfnamefont {M.~C.}\ \bibnamefont
  {Marchetti}},\ }\bibfield  {title} {\bibinfo {title} {Athermal phase
  separation of self-propelled particles with no alignment},\ }\href
  {https://doi.org/https://doi.org/10.1103/PhysRevLett.108.235702} {\bibfield
  {journal} {\bibinfo  {journal} {Physical Review Letters}\ }\textbf {\bibinfo
  {volume} {108}},\ \bibinfo {pages} {235702} (\bibinfo {year}
  {2012})}\BibitemShut {NoStop}%
\bibitem [{\citenamefont {Marchetti}\ \emph {et~al.}(2013)\citenamefont
  {Marchetti}, \citenamefont {Joanny}, \citenamefont {Ramaswamy}, \citenamefont
  {Liverpool}, \citenamefont {Prost}, \citenamefont {Rao},\ and\ \citenamefont
  {Simha}}]{RevModPhys.85.1143}%
  \BibitemOpen
  \bibfield  {author} {\bibinfo {author} {\bibfnamefont {M.~C.}\ \bibnamefont
  {Marchetti}}, \bibinfo {author} {\bibfnamefont {J.~F.}\ \bibnamefont
  {Joanny}}, \bibinfo {author} {\bibfnamefont {S.}~\bibnamefont {Ramaswamy}},
  \bibinfo {author} {\bibfnamefont {T.~B.}\ \bibnamefont {Liverpool}}, \bibinfo
  {author} {\bibfnamefont {J.}~\bibnamefont {Prost}}, \bibinfo {author}
  {\bibfnamefont {M.}~\bibnamefont {Rao}},\ and\ \bibinfo {author}
  {\bibfnamefont {R.~A.}\ \bibnamefont {Simha}},\ }\bibfield  {title} {\bibinfo
  {title} {Hydrodynamics of soft active matter},\ }\href
  {https://doi.org/10.1103/RevModPhys.85.1143} {\bibfield  {journal} {\bibinfo
  {journal} {Rev. Mod. Phys.}\ }\textbf {\bibinfo {volume} {85}},\ \bibinfo
  {pages} {1143} (\bibinfo {year} {2013})}\BibitemShut {NoStop}%
\bibitem [{\citenamefont {Bechinger}\ \emph {et~al.}(2016)\citenamefont
  {Bechinger}, \citenamefont {Di~Leonardo}, \citenamefont {L\"owen},
  \citenamefont {Reichhardt}, \citenamefont {Volpe},\ and\ \citenamefont
  {Volpe}}]{RevModPhys.88.045006}%
  \BibitemOpen
  \bibfield  {author} {\bibinfo {author} {\bibfnamefont {C.}~\bibnamefont
  {Bechinger}}, \bibinfo {author} {\bibfnamefont {R.}~\bibnamefont
  {Di~Leonardo}}, \bibinfo {author} {\bibfnamefont {H.}~\bibnamefont
  {L\"owen}}, \bibinfo {author} {\bibfnamefont {C.}~\bibnamefont {Reichhardt}},
  \bibinfo {author} {\bibfnamefont {G.}~\bibnamefont {Volpe}},\ and\ \bibinfo
  {author} {\bibfnamefont {G.}~\bibnamefont {Volpe}},\ }\bibfield  {title}
  {\bibinfo {title} {Active particles in complex and crowded environments},\
  }\href {https://doi.org/10.1103/RevModPhys.88.045006} {\bibfield  {journal}
  {\bibinfo  {journal} {Rev. Mod. Phys.}\ }\textbf {\bibinfo {volume} {88}},\
  \bibinfo {pages} {045006} (\bibinfo {year} {2016})}\BibitemShut {NoStop}%
\bibitem [{\citenamefont {Cates}\ and\ \citenamefont
  {Tailleur}(2015)}]{cates2015motility}%
  \BibitemOpen
  \bibfield  {author} {\bibinfo {author} {\bibfnamefont {M.~E.}\ \bibnamefont
  {Cates}}\ and\ \bibinfo {author} {\bibfnamefont {J.}~\bibnamefont
  {Tailleur}},\ }\bibfield  {title} {\bibinfo {title} {Motility-induced phase
  separation},\ }\href
  {https://doi.org/https://doi.org/10.1146/annurev-conmatphys-031214-014710}
  {\bibfield  {journal} {\bibinfo  {journal} {Annu. Rev. Condens. Matter
  Phys.}\ }\textbf {\bibinfo {volume} {6}},\ \bibinfo {pages} {219} (\bibinfo
  {year} {2015})}\BibitemShut {NoStop}%
\bibitem [{\citenamefont {Palacci}\ \emph {et~al.}(2013)\citenamefont
  {Palacci}, \citenamefont {Sacanna}, \citenamefont {Steinberg}, \citenamefont
  {Pine},\ and\ \citenamefont {Chaikin}}]{palacci2013living}%
  \BibitemOpen
  \bibfield  {author} {\bibinfo {author} {\bibfnamefont {J.}~\bibnamefont
  {Palacci}}, \bibinfo {author} {\bibfnamefont {S.}~\bibnamefont {Sacanna}},
  \bibinfo {author} {\bibfnamefont {A.~P.}\ \bibnamefont {Steinberg}}, \bibinfo
  {author} {\bibfnamefont {D.~J.}\ \bibnamefont {Pine}},\ and\ \bibinfo
  {author} {\bibfnamefont {P.~M.}\ \bibnamefont {Chaikin}},\ }\bibfield
  {title} {\bibinfo {title} {Living crystals of light-activated colloidal
  surfers},\ }\href {https://doi.org/10.1126/science.1230020} {\bibfield
  {journal} {\bibinfo  {journal} {Science}\ }\textbf {\bibinfo {volume}
  {339}},\ \bibinfo {pages} {936} (\bibinfo {year} {2013})}\BibitemShut
  {NoStop}%
\bibitem [{\citenamefont {Bialk{\'e}}\ \emph {et~al.}(2012)\citenamefont
  {Bialk{\'e}}, \citenamefont {Speck},\ and\ \citenamefont
  {L{\"o}wen}}]{bialke2012crystallization}%
  \BibitemOpen
  \bibfield  {author} {\bibinfo {author} {\bibfnamefont {J.}~\bibnamefont
  {Bialk{\'e}}}, \bibinfo {author} {\bibfnamefont {T.}~\bibnamefont {Speck}},\
  and\ \bibinfo {author} {\bibfnamefont {H.}~\bibnamefont {L{\"o}wen}},\
  }\bibfield  {title} {\bibinfo {title} {Crystallization in a dense suspension
  of self-propelled particles},\ }\href
  {https://doi.org/https://doi.org/10.1103/PhysRevLett.108.168301} {\bibfield
  {journal} {\bibinfo  {journal} {Physical Review Letters}\ }\textbf {\bibinfo
  {volume} {108}},\ \bibinfo {pages} {168301} (\bibinfo {year}
  {2012})}\BibitemShut {NoStop}%
\bibitem [{\citenamefont {Redner}\ \emph {et~al.}(2013)\citenamefont {Redner},
  \citenamefont {Hagan},\ and\ \citenamefont {Baskaran}}]{redner2013structure}%
  \BibitemOpen
  \bibfield  {author} {\bibinfo {author} {\bibfnamefont {G.~S.}\ \bibnamefont
  {Redner}}, \bibinfo {author} {\bibfnamefont {M.~F.}\ \bibnamefont {Hagan}},\
  and\ \bibinfo {author} {\bibfnamefont {A.}~\bibnamefont {Baskaran}},\
  }\bibfield  {title} {\bibinfo {title} {Structure and dynamics of a
  phase-separating active colloidal fluid},\ }\href
  {https://doi.org/https://doi.org/10.1103/PhysRevLett.110.055701} {\bibfield
  {journal} {\bibinfo  {journal} {Physical Review Letters}\ }\textbf {\bibinfo
  {volume} {110}},\ \bibinfo {pages} {055701} (\bibinfo {year}
  {2013})}\BibitemShut {NoStop}%
\bibitem [{\citenamefont {Digregorio}\ \emph {et~al.}(2018)\citenamefont
  {Digregorio}, \citenamefont {Levis}, \citenamefont {Suma}, \citenamefont
  {Cugliandolo}, \citenamefont {Gonnella},\ and\ \citenamefont
  {Pagonabarraga}}]{digregorio2018full}%
  \BibitemOpen
  \bibfield  {author} {\bibinfo {author} {\bibfnamefont {P.}~\bibnamefont
  {Digregorio}}, \bibinfo {author} {\bibfnamefont {D.}~\bibnamefont {Levis}},
  \bibinfo {author} {\bibfnamefont {A.}~\bibnamefont {Suma}}, \bibinfo {author}
  {\bibfnamefont {L.~F.}\ \bibnamefont {Cugliandolo}}, \bibinfo {author}
  {\bibfnamefont {G.}~\bibnamefont {Gonnella}},\ and\ \bibinfo {author}
  {\bibfnamefont {I.}~\bibnamefont {Pagonabarraga}},\ }\bibfield  {title}
  {\bibinfo {title} {Full phase diagram of active brownian disks: From melting
  to motility-induced phase separation},\ }\href
  {https://doi.org/https://doi.org/10.1103/PhysRevLett.121.098003} {\bibfield
  {journal} {\bibinfo  {journal} {Physical Review Letters}\ }\textbf {\bibinfo
  {volume} {121}},\ \bibinfo {pages} {098003} (\bibinfo {year}
  {2018})}\BibitemShut {NoStop}%
\bibitem [{\citenamefont {Omar}\ \emph {et~al.}(2021)\citenamefont {Omar},
  \citenamefont {Klymko}, \citenamefont {GrandPre},\ and\ \citenamefont
  {Geissler}}]{omar2021phase}%
  \BibitemOpen
  \bibfield  {author} {\bibinfo {author} {\bibfnamefont {A.~K.}\ \bibnamefont
  {Omar}}, \bibinfo {author} {\bibfnamefont {K.}~\bibnamefont {Klymko}},
  \bibinfo {author} {\bibfnamefont {T.}~\bibnamefont {GrandPre}},\ and\
  \bibinfo {author} {\bibfnamefont {P.~L.}\ \bibnamefont {Geissler}},\
  }\bibfield  {title} {\bibinfo {title} {Phase diagram of active brownian
  spheres: Crystallization and the metastability of motility-induced phase
  separation},\ }\href
  {https://doi.org/https://doi.org/10.1103/PhysRevLett.126.188002} {\bibfield
  {journal} {\bibinfo  {journal} {Physical Review Letters}\ }\textbf {\bibinfo
  {volume} {126}},\ \bibinfo {pages} {188002} (\bibinfo {year}
  {2021})}\BibitemShut {NoStop}%
\bibitem [{\citenamefont {Henkes}\ \emph {et~al.}(2011)\citenamefont {Henkes},
  \citenamefont {Fily},\ and\ \citenamefont {Marchetti}}]{Henkes2011}%
  \BibitemOpen
  \bibfield  {author} {\bibinfo {author} {\bibfnamefont {S.}~\bibnamefont
  {Henkes}}, \bibinfo {author} {\bibfnamefont {Y.}~\bibnamefont {Fily}},\ and\
  \bibinfo {author} {\bibfnamefont {M.~C.}\ \bibnamefont {Marchetti}},\
  }\bibfield  {title} {\bibinfo {title} {{Active jamming: Self-propelled soft
  particles at high density}},\ }\href
  {https://doi.org/10.1103/PhysRevE.84.040301} {\bibfield  {journal} {\bibinfo
  {journal} {Physical Review E}\ }\textbf {\bibinfo {volume} {84}},\ \bibinfo
  {pages} {040301} (\bibinfo {year} {2011})}\BibitemShut {NoStop}%
\bibitem [{\citenamefont {Henkes}\ \emph {et~al.}(2020)\citenamefont {Henkes},
  \citenamefont {Kostanjevec}, \citenamefont {Collinson}, \citenamefont
  {Sknepnek},\ and\ \citenamefont {Bertin}}]{henkes2020dense}%
  \BibitemOpen
  \bibfield  {author} {\bibinfo {author} {\bibfnamefont {S.}~\bibnamefont
  {Henkes}}, \bibinfo {author} {\bibfnamefont {K.}~\bibnamefont {Kostanjevec}},
  \bibinfo {author} {\bibfnamefont {J.~M.}\ \bibnamefont {Collinson}}, \bibinfo
  {author} {\bibfnamefont {R.}~\bibnamefont {Sknepnek}},\ and\ \bibinfo
  {author} {\bibfnamefont {E.}~\bibnamefont {Bertin}},\ }\bibfield  {title}
  {\bibinfo {title} {Dense active matter model of motion patterns in confluent
  cell monolayers},\ }\href
  {https://doi.org/https://doi.org/10.1038/s41467-020-15164-5} {\bibfield
  {journal} {\bibinfo  {journal} {Nature Communications}\ }\textbf {\bibinfo
  {volume} {11}},\ \bibinfo {pages} {1} (\bibinfo {year} {2020})}\BibitemShut
  {NoStop}%
\bibitem [{\citenamefont {Bi}\ \emph {et~al.}(2016)\citenamefont {Bi},
  \citenamefont {Yang}, \citenamefont {Marchetti},\ and\ \citenamefont
  {Manning}}]{Bi2016}%
  \BibitemOpen
  \bibfield  {author} {\bibinfo {author} {\bibfnamefont {D.}~\bibnamefont
  {Bi}}, \bibinfo {author} {\bibfnamefont {X.}~\bibnamefont {Yang}}, \bibinfo
  {author} {\bibfnamefont {M.~C.}\ \bibnamefont {Marchetti}},\ and\ \bibinfo
  {author} {\bibfnamefont {M.~L.}\ \bibnamefont {Manning}},\ }\bibfield
  {title} {\bibinfo {title} {{Motility-Driven Glass and Jamming Transitions in
  Biological Tissues}},\ }\href {https://doi.org/10.1103/PhysRevX.6.021011}
  {\bibfield  {journal} {\bibinfo  {journal} {Physical Review X}\ }\textbf
  {\bibinfo {volume} {6}},\ \bibinfo {pages} {021011} (\bibinfo {year}
  {2016})}\BibitemShut {NoStop}%
\bibitem [{\citenamefont {Garcia}\ \emph {et~al.}(2015)\citenamefont {Garcia},
  \citenamefont {Hannezo}, \citenamefont {Elgeti}, \citenamefont {Joanny},
  \citenamefont {Silberzan},\ and\ \citenamefont {Gov}}]{garcia2015physics}%
  \BibitemOpen
  \bibfield  {author} {\bibinfo {author} {\bibfnamefont {S.}~\bibnamefont
  {Garcia}}, \bibinfo {author} {\bibfnamefont {E.}~\bibnamefont {Hannezo}},
  \bibinfo {author} {\bibfnamefont {J.}~\bibnamefont {Elgeti}}, \bibinfo
  {author} {\bibfnamefont {J.-F.}\ \bibnamefont {Joanny}}, \bibinfo {author}
  {\bibfnamefont {P.}~\bibnamefont {Silberzan}},\ and\ \bibinfo {author}
  {\bibfnamefont {N.~S.}\ \bibnamefont {Gov}},\ }\bibfield  {title} {\bibinfo
  {title} {Physics of active jamming during collective cellular motion in a
  monolayer},\ }\href {https://doi.org/https://doi.org/10.1073/pnas.1510973112}
  {\bibfield  {journal} {\bibinfo  {journal} {Proceedings of the National
  Academy of Sciences}\ }\textbf {\bibinfo {volume} {112}},\ \bibinfo {pages}
  {15314} (\bibinfo {year} {2015})}\BibitemShut {NoStop}%
\bibitem [{\citenamefont {Giavazzi}\ \emph {et~al.}(2018)\citenamefont
  {Giavazzi}, \citenamefont {Paoluzzi}, \citenamefont {Macchi}, \citenamefont
  {Bi}, \citenamefont {Scita}, \citenamefont {Manning}, \citenamefont
  {Cerbino},\ and\ \citenamefont {Marchetti}}]{Giavazzi2018}%
  \BibitemOpen
  \bibfield  {author} {\bibinfo {author} {\bibfnamefont {F.}~\bibnamefont
  {Giavazzi}}, \bibinfo {author} {\bibfnamefont {M.}~\bibnamefont {Paoluzzi}},
  \bibinfo {author} {\bibfnamefont {M.}~\bibnamefont {Macchi}}, \bibinfo
  {author} {\bibfnamefont {D.}~\bibnamefont {Bi}}, \bibinfo {author}
  {\bibfnamefont {G.}~\bibnamefont {Scita}}, \bibinfo {author} {\bibfnamefont
  {M.~L.}\ \bibnamefont {Manning}}, \bibinfo {author} {\bibfnamefont
  {R.}~\bibnamefont {Cerbino}},\ and\ \bibinfo {author} {\bibfnamefont {M.~C.}\
  \bibnamefont {Marchetti}},\ }\bibfield  {title} {\bibinfo {title} {{Flocking
  transitions in confluent tissues}},\ }\href
  {https://doi.org/10.1039/c8sm00126j} {\bibfield  {journal} {\bibinfo
  {journal} {Soft Matter}\ }\textbf {\bibinfo {volume} {14}},\ \bibinfo {pages}
  {3471} (\bibinfo {year} {2018})}\BibitemShut {NoStop}%
\bibitem [{\citenamefont {Mongera}\ \emph {et~al.}(2018)\citenamefont
  {Mongera}, \citenamefont {Rowghanian}, \citenamefont {Gustafson},
  \citenamefont {Shelton}, \citenamefont {Kealhofer}, \citenamefont {Carn},
  \citenamefont {Serwane}, \citenamefont {Lucio}, \citenamefont {Giammona},\
  and\ \citenamefont {Camp{\`a}s}}]{mongera2018fluid}%
  \BibitemOpen
  \bibfield  {author} {\bibinfo {author} {\bibfnamefont {A.}~\bibnamefont
  {Mongera}}, \bibinfo {author} {\bibfnamefont {P.}~\bibnamefont {Rowghanian}},
  \bibinfo {author} {\bibfnamefont {H.~J.}\ \bibnamefont {Gustafson}}, \bibinfo
  {author} {\bibfnamefont {E.}~\bibnamefont {Shelton}}, \bibinfo {author}
  {\bibfnamefont {D.~A.}\ \bibnamefont {Kealhofer}}, \bibinfo {author}
  {\bibfnamefont {E.~K.}\ \bibnamefont {Carn}}, \bibinfo {author}
  {\bibfnamefont {F.}~\bibnamefont {Serwane}}, \bibinfo {author} {\bibfnamefont
  {A.~A.}\ \bibnamefont {Lucio}}, \bibinfo {author} {\bibfnamefont
  {J.}~\bibnamefont {Giammona}},\ and\ \bibinfo {author} {\bibfnamefont
  {O.}~\bibnamefont {Camp{\`a}s}},\ }\bibfield  {title} {\bibinfo {title} {A
  fluid-to-solid jamming transition underlies vertebrate body axis
  elongation},\ }\href
  {https://doi.org/https://doi.org/10.1038/s41586-018-0479-2} {\bibfield
  {journal} {\bibinfo  {journal} {Nature}\ }\textbf {\bibinfo {volume} {561}},\
  \bibinfo {pages} {401} (\bibinfo {year} {2018})}\BibitemShut {NoStop}%
\bibitem [{\citenamefont {Pasupalak}\ \emph {et~al.}(2020)\citenamefont
  {Pasupalak}, \citenamefont {Yan-Wei}, \citenamefont {Ni},\ and\ \citenamefont
  {{Pica Ciamarra}}}]{D0SM00109K}%
  \BibitemOpen
  \bibfield  {author} {\bibinfo {author} {\bibfnamefont {A.}~\bibnamefont
  {Pasupalak}}, \bibinfo {author} {\bibfnamefont {L.}~\bibnamefont {Yan-Wei}},
  \bibinfo {author} {\bibfnamefont {R.}~\bibnamefont {Ni}},\ and\ \bibinfo
  {author} {\bibfnamefont {M.}~\bibnamefont {{Pica Ciamarra}}},\ }\bibfield
  {title} {\bibinfo {title} {{Hexatic phase in a model of active biological
  tissues}},\ }\href {https://doi.org/10.1039/D0SM00109K} {\bibfield  {journal}
  {\bibinfo  {journal} {Soft Matter}\ }\textbf {\bibinfo {volume} {16}},\
  \bibinfo {pages} {3914} (\bibinfo {year} {2020})}\BibitemShut {NoStop}%
\bibitem [{\citenamefont {Li}\ \emph {et~al.}(2021)\citenamefont {Li},
  \citenamefont {Wei}, \citenamefont {Paoluzzi},\ and\ \citenamefont
  {Ciamarra}}]{Li2021}%
  \BibitemOpen
  \bibfield  {author} {\bibinfo {author} {\bibfnamefont {Y.-W.}\ \bibnamefont
  {Li}}, \bibinfo {author} {\bibfnamefont {L.~L.~Y.}\ \bibnamefont {Wei}},
  \bibinfo {author} {\bibfnamefont {M.}~\bibnamefont {Paoluzzi}},\ and\
  \bibinfo {author} {\bibfnamefont {M.~P.}\ \bibnamefont {Ciamarra}},\
  }\bibfield  {title} {\bibinfo {title} {{Softness, anomalous dynamics, and
  fractal-like energy landscape in model cell tissues}},\ }\href
  {https://doi.org/10.1103/PhysRevE.103.022607} {\bibfield  {journal} {\bibinfo
   {journal} {Physical Review E}\ }\textbf {\bibinfo {volume} {103}},\ \bibinfo
  {pages} {022607} (\bibinfo {year} {2021})}\BibitemShut {NoStop}%
\bibitem [{\citenamefont {Lawson-Keister}\ and\ \citenamefont
  {Manning}(2021)}]{Lawson-Keister2021}%
  \BibitemOpen
  \bibfield  {author} {\bibinfo {author} {\bibfnamefont {E.}~\bibnamefont
  {Lawson-Keister}}\ and\ \bibinfo {author} {\bibfnamefont {M.~L.}\
  \bibnamefont {Manning}},\ }\bibfield  {title} {\bibinfo {title} {{Jamming and
  arrest of cell motion in biological tissues}},\ }\href
  {https://doi.org/10.1016/j.ceb.2021.07.011} {\bibfield  {journal} {\bibinfo
  {journal} {Current Opinion in Cell Biology}\ }\textbf {\bibinfo {volume}
  {72}},\ \bibinfo {pages} {146} (\bibinfo {year} {2021})}\BibitemShut
  {NoStop}%
\bibitem [{\citenamefont {Parry}\ \emph {et~al.}(2014)\citenamefont {Parry},
  \citenamefont {Surovtsev}, \citenamefont {Cabeen}, \citenamefont {O’Hern},
  \citenamefont {Dufresne},\ and\ \citenamefont
  {Jacobs-Wagner}}]{parry2014bacterial}%
  \BibitemOpen
  \bibfield  {author} {\bibinfo {author} {\bibfnamefont {B.~R.}\ \bibnamefont
  {Parry}}, \bibinfo {author} {\bibfnamefont {I.~V.}\ \bibnamefont
  {Surovtsev}}, \bibinfo {author} {\bibfnamefont {M.~T.}\ \bibnamefont
  {Cabeen}}, \bibinfo {author} {\bibfnamefont {C.~S.}\ \bibnamefont
  {O’Hern}}, \bibinfo {author} {\bibfnamefont {E.~R.}\ \bibnamefont
  {Dufresne}},\ and\ \bibinfo {author} {\bibfnamefont {C.}~\bibnamefont
  {Jacobs-Wagner}},\ }\bibfield  {title} {\bibinfo {title} {The bacterial
  cytoplasm has glass-like properties and is fluidized by metabolic activity},\
  }\href {https://doi.org/https://doi.org/10.1016/j.cell.2013.11.028}
  {\bibfield  {journal} {\bibinfo  {journal} {Cell}\ }\textbf {\bibinfo
  {volume} {156}},\ \bibinfo {pages} {183} (\bibinfo {year}
  {2014})}\BibitemShut {NoStop}%
\bibitem [{\citenamefont {Delarue}\ \emph {et~al.}(2016)\citenamefont
  {Delarue}, \citenamefont {Hartung}, \citenamefont {Schreck}, \citenamefont
  {Gniewek}, \citenamefont {Hu}, \citenamefont {Herminghaus},\ and\
  \citenamefont {Hallatschek}}]{Delarue2016}%
  \BibitemOpen
  \bibfield  {author} {\bibinfo {author} {\bibfnamefont {M.}~\bibnamefont
  {Delarue}}, \bibinfo {author} {\bibfnamefont {J.}~\bibnamefont {Hartung}},
  \bibinfo {author} {\bibfnamefont {C.~F.}\ \bibnamefont {Schreck}}, \bibinfo
  {author} {\bibfnamefont {P.}~\bibnamefont {Gniewek}}, \bibinfo {author}
  {\bibfnamefont {L.}~\bibnamefont {Hu}}, \bibinfo {author} {\bibfnamefont
  {S.}~\bibnamefont {Herminghaus}},\ and\ \bibinfo {author} {\bibfnamefont
  {O.}~\bibnamefont {Hallatschek}},\ }\bibfield  {title} {\bibinfo {title}
  {{Self-driven jamming in growing microbial populations}},\ }\href
  {https://doi.org/10.1038/nphys3741} {\bibfield  {journal} {\bibinfo
  {journal} {Nature Physics}\ }\textbf {\bibinfo {volume} {12}},\ \bibinfo
  {pages} {762} (\bibinfo {year} {2016})}\BibitemShut {NoStop}%
\bibitem [{\citenamefont {Yang}\ \emph {et~al.}(2019)\citenamefont {Yang},
  \citenamefont {Arratia}, \citenamefont {Patteson},\ and\ \citenamefont
  {Gopinath}}]{yang2019quenching}%
  \BibitemOpen
  \bibfield  {author} {\bibinfo {author} {\bibfnamefont {J.}~\bibnamefont
  {Yang}}, \bibinfo {author} {\bibfnamefont {P.~E.}\ \bibnamefont {Arratia}},
  \bibinfo {author} {\bibfnamefont {A.~E.}\ \bibnamefont {Patteson}},\ and\
  \bibinfo {author} {\bibfnamefont {A.}~\bibnamefont {Gopinath}},\ }\bibfield
  {title} {\bibinfo {title} {Quenching active swarms: effects of light exposure
  on collective motility in swarming serratia marcescens},\ }\href
  {https://doi.org/https://doi.org/10.1098/rsif.2018.0960} {\bibfield
  {journal} {\bibinfo  {journal} {Journal of the Royal Society Interface}\
  }\textbf {\bibinfo {volume} {16}},\ \bibinfo {pages} {20180960} (\bibinfo
  {year} {2019})}\BibitemShut {NoStop}%
\bibitem [{\citenamefont {Mandal}\ \emph {et~al.}(2020)\citenamefont {Mandal},
  \citenamefont {Bhuyan}, \citenamefont {Chaudhuri}, \citenamefont {Dasgupta},\
  and\ \citenamefont {Rao}}]{mandal2020extreme}%
  \BibitemOpen
  \bibfield  {author} {\bibinfo {author} {\bibfnamefont {R.}~\bibnamefont
  {Mandal}}, \bibinfo {author} {\bibfnamefont {P.~J.}\ \bibnamefont {Bhuyan}},
  \bibinfo {author} {\bibfnamefont {P.}~\bibnamefont {Chaudhuri}}, \bibinfo
  {author} {\bibfnamefont {C.}~\bibnamefont {Dasgupta}},\ and\ \bibinfo
  {author} {\bibfnamefont {M.}~\bibnamefont {Rao}},\ }\bibfield  {title}
  {\bibinfo {title} {Extreme active matter at high densities},\ }\href
  {https://doi.org/https://doi.org/10.1038/s41467-020-16130-x} {\bibfield
  {journal} {\bibinfo  {journal} {Nature Communications}\ }\textbf {\bibinfo
  {volume} {11}},\ \bibinfo {pages} {1} (\bibinfo {year} {2020})}\BibitemShut
  {NoStop}%
\bibitem [{\citenamefont {Mandal}\ and\ \citenamefont
  {Sollich}(2020)}]{mandal2020multiple}%
  \BibitemOpen
  \bibfield  {author} {\bibinfo {author} {\bibfnamefont {R.}~\bibnamefont
  {Mandal}}\ and\ \bibinfo {author} {\bibfnamefont {P.}~\bibnamefont
  {Sollich}},\ }\bibfield  {title} {\bibinfo {title} {Multiple types of aging
  in active glasses},\ }\href
  {https://doi.org/https://doi.org/10.1103/PhysRevLett.125.218001} {\bibfield
  {journal} {\bibinfo  {journal} {Physical Review Letters}\ }\textbf {\bibinfo
  {volume} {125}},\ \bibinfo {pages} {218001} (\bibinfo {year}
  {2020})}\BibitemShut {NoStop}%
\bibitem [{\citenamefont {Paoluzzi}\ \emph {et~al.}(2021)\citenamefont
  {Paoluzzi}, \citenamefont {Levis},\ and\ \citenamefont
  {Pagonabarraga}}]{Paoluzzi2021}%
  \BibitemOpen
  \bibfield  {author} {\bibinfo {author} {\bibfnamefont {M.}~\bibnamefont
  {Paoluzzi}}, \bibinfo {author} {\bibfnamefont {D.}~\bibnamefont {Levis}},\
  and\ \bibinfo {author} {\bibfnamefont {I.}~\bibnamefont {Pagonabarraga}},\
  }\href {https://doi.org/10.48550/ARXIV.2109.14948} {\bibinfo {title} {From
  motility-induced phase-separation to glassiness in dense active matter}}
  (\bibinfo {year} {2021}),\ \Eprint {https://arxiv.org/abs/2109.14948}
  {arXiv:2109.14948 [cond-mat.soft]} \BibitemShut {NoStop}%
\bibitem [{\citenamefont {Nie}\ \emph {et~al.}(2020)\citenamefont {Nie},
  \citenamefont {Chattoraj}, \citenamefont {Piscitelli}, \citenamefont {Doyle},
  \citenamefont {Ni},\ and\ \citenamefont {Ciamarra}}]{nie2020stability}%
  \BibitemOpen
  \bibfield  {author} {\bibinfo {author} {\bibfnamefont {P.}~\bibnamefont
  {Nie}}, \bibinfo {author} {\bibfnamefont {J.}~\bibnamefont {Chattoraj}},
  \bibinfo {author} {\bibfnamefont {A.}~\bibnamefont {Piscitelli}}, \bibinfo
  {author} {\bibfnamefont {P.}~\bibnamefont {Doyle}}, \bibinfo {author}
  {\bibfnamefont {R.}~\bibnamefont {Ni}},\ and\ \bibinfo {author}
  {\bibfnamefont {M.~P.}\ \bibnamefont {Ciamarra}},\ }\bibfield  {title}
  {\bibinfo {title} {Stability phase diagram of active brownian particles},\
  }\href {https://doi.org/https://doi.org/10.1103/PhysRevResearch.2.023010}
  {\bibfield  {journal} {\bibinfo  {journal} {Physical Review Research}\
  }\textbf {\bibinfo {volume} {2}},\ \bibinfo {pages} {023010} (\bibinfo {year}
  {2020})}\BibitemShut {NoStop}%
\bibitem [{\citenamefont {Fily}\ \emph {et~al.}(2014)\citenamefont {Fily},
  \citenamefont {Henkes},\ and\ \citenamefont {Marchetti}}]{fily2014freezing}%
  \BibitemOpen
  \bibfield  {author} {\bibinfo {author} {\bibfnamefont {Y.}~\bibnamefont
  {Fily}}, \bibinfo {author} {\bibfnamefont {S.}~\bibnamefont {Henkes}},\ and\
  \bibinfo {author} {\bibfnamefont {M.~C.}\ \bibnamefont {Marchetti}},\
  }\bibfield  {title} {\bibinfo {title} {Freezing and phase separation of
  self-propelled disks},\ }\href
  {https://doi.org/https://doi.org/10.1039/C3SM52469H} {\bibfield  {journal}
  {\bibinfo  {journal} {Soft Matter}\ }\textbf {\bibinfo {volume} {10}},\
  \bibinfo {pages} {2132} (\bibinfo {year} {2014})}\BibitemShut {NoStop}%
\bibitem [{\citenamefont {Keta}\ \emph {et~al.}(2022)\citenamefont {Keta},
  \citenamefont {Jack},\ and\ \citenamefont {Berthier}}]{keta2022disordered}%
  \BibitemOpen
  \bibfield  {author} {\bibinfo {author} {\bibfnamefont {Y.-E.}\ \bibnamefont
  {Keta}}, \bibinfo {author} {\bibfnamefont {R.~L.}\ \bibnamefont {Jack}},\
  and\ \bibinfo {author} {\bibfnamefont {L.}~\bibnamefont {Berthier}},\
  }\href@noop {} {\bibinfo {title} {{Disordered collective motion in dense
  assemblies of persistent particles}}} (\bibinfo {year} {2022}),\ \Eprint
  {https://arxiv.org/abs/2201.04902} {arXiv:2201.04902 [cond-mat.soft]}
  \BibitemShut {NoStop}%
\bibitem [{\citenamefont {Merrigan}\ \emph {et~al.}(2020)\citenamefont
  {Merrigan}, \citenamefont {Ramola}, \citenamefont {Chatterjee}, \citenamefont
  {Segall}, \citenamefont {Shokef},\ and\ \citenamefont
  {Chakraborty}}]{merrigan2020arrested}%
  \BibitemOpen
  \bibfield  {author} {\bibinfo {author} {\bibfnamefont {C.}~\bibnamefont
  {Merrigan}}, \bibinfo {author} {\bibfnamefont {K.}~\bibnamefont {Ramola}},
  \bibinfo {author} {\bibfnamefont {R.}~\bibnamefont {Chatterjee}}, \bibinfo
  {author} {\bibfnamefont {N.}~\bibnamefont {Segall}}, \bibinfo {author}
  {\bibfnamefont {Y.}~\bibnamefont {Shokef}},\ and\ \bibinfo {author}
  {\bibfnamefont {B.}~\bibnamefont {Chakraborty}},\ }\bibfield  {title}
  {\bibinfo {title} {Arrested states in persistent active matter: Gelation
  without attraction},\ }\href
  {https://doi.org/https://doi.org/10.1103/PhysRevResearch.2.013260} {\bibfield
   {journal} {\bibinfo  {journal} {Physical Review Research}\ }\textbf
  {\bibinfo {volume} {2}},\ \bibinfo {pages} {013260} (\bibinfo {year}
  {2020})}\BibitemShut {NoStop}%
\bibitem [{\citenamefont {Liao}\ and\ \citenamefont
  {Xu}(2018)}]{liao2018criticality}%
  \BibitemOpen
  \bibfield  {author} {\bibinfo {author} {\bibfnamefont {Q.}~\bibnamefont
  {Liao}}\ and\ \bibinfo {author} {\bibfnamefont {N.}~\bibnamefont {Xu}},\
  }\bibfield  {title} {\bibinfo {title} {Criticality of the zero-temperature
  jamming transition probed by self-propelled particles},\ }\href
  {https://doi.org/https://doi.org/10.1039/C7SM01909B} {\bibfield  {journal}
  {\bibinfo  {journal} {Soft Matter}\ }\textbf {\bibinfo {volume} {14}},\
  \bibinfo {pages} {853} (\bibinfo {year} {2018})}\BibitemShut {NoStop}%
\bibitem [{\citenamefont {Reichhardt}\ and\ \citenamefont {{Olson
  Reichhardt}}(2014)}]{Reichhardt2014}%
  \BibitemOpen
  \bibfield  {author} {\bibinfo {author} {\bibfnamefont {C.}~\bibnamefont
  {Reichhardt}}\ and\ \bibinfo {author} {\bibfnamefont {C.~J.}\ \bibnamefont
  {{Olson Reichhardt}}},\ }\bibfield  {title} {\bibinfo {title} {{Absorbing
  phase transitions and dynamic freezing in running active matter systems}},\
  }\bibfield  {journal} {\bibinfo  {journal} {Soft Matter}\ }\textbf {\bibinfo
  {volume} {10}},\ \href {https://doi.org/10.1039/c4sm01273a}
  {10.1039/c4sm01273a} (\bibinfo {year} {2014})\BibitemShut {NoStop}%
\bibitem [{\citenamefont {Br{\"u}ning}\ \emph {et~al.}(2008)\citenamefont
  {Br{\"u}ning}, \citenamefont {St-Onge}, \citenamefont {Patterson},\ and\
  \citenamefont {Kob}}]{bruning2008glass}%
  \BibitemOpen
  \bibfield  {author} {\bibinfo {author} {\bibfnamefont {R.}~\bibnamefont
  {Br{\"u}ning}}, \bibinfo {author} {\bibfnamefont {D.~A.}\ \bibnamefont
  {St-Onge}}, \bibinfo {author} {\bibfnamefont {S.}~\bibnamefont {Patterson}},\
  and\ \bibinfo {author} {\bibfnamefont {W.}~\bibnamefont {Kob}},\ }\bibfield
  {title} {\bibinfo {title} {Glass transitions in one-, two-, three-, and
  four-dimensional binary lennard-jones systems},\ }\href
  {https://doi.org/https://doi.org/10.1088/0953-8984/21/3/035117} {\bibfield
  {journal} {\bibinfo  {journal} {Journal of Physics: Condensed Matter}\
  }\textbf {\bibinfo {volume} {21}},\ \bibinfo {pages} {035117} (\bibinfo
  {year} {2008})}\BibitemShut {NoStop}%
\bibitem [{Note1()}]{Note1}%
  \BibitemOpen
  \bibinfo {note} {$\langle a \rangle = \protect \frac {\pi }{4}[0.64
  d_{\protect \rm AA}^2+0.35 d_{\protect \rm BB}^2])$}\BibitemShut {NoStop}%
\bibitem [{\citenamefont {Chaudhuri}\ \emph {et~al.}(2010)\citenamefont
  {Chaudhuri}, \citenamefont {Berthier},\ and\ \citenamefont
  {Sastry}}]{Chaudhuri2010}%
  \BibitemOpen
  \bibfield  {author} {\bibinfo {author} {\bibfnamefont {P.}~\bibnamefont
  {Chaudhuri}}, \bibinfo {author} {\bibfnamefont {L.}~\bibnamefont
  {Berthier}},\ and\ \bibinfo {author} {\bibfnamefont {S.}~\bibnamefont
  {Sastry}},\ }\bibfield  {title} {\bibinfo {title} {{Jamming Transitions in
  Amorphous Packings of Frictionless Spheres Occur over a Continuous Range of
  Volume Fractions}},\ }\href {https://doi.org/10.1103/PhysRevLett.104.165701}
  {\bibfield  {journal} {\bibinfo  {journal} {Physical Review Letters}\
  }\textbf {\bibinfo {volume} {104}},\ \bibinfo {pages} {165701} (\bibinfo
  {year} {2010})}\BibitemShut {NoStop}%
\bibitem [{\citenamefont {{Pica Ciamarra}}\ \emph {et~al.}(2010)\citenamefont
  {{Pica Ciamarra}}, \citenamefont {Coniglio},\ and\ \citenamefont {{De
  Candia}}}]{PicaCiamarra2010}%
  \BibitemOpen
  \bibfield  {author} {\bibinfo {author} {\bibfnamefont {M.}~\bibnamefont
  {{Pica Ciamarra}}}, \bibinfo {author} {\bibfnamefont {A.}~\bibnamefont
  {Coniglio}},\ and\ \bibinfo {author} {\bibfnamefont {A.}~\bibnamefont {{De
  Candia}}},\ }\bibfield  {title} {\bibinfo {title} {{Disordered jammed
  packings of frictionless spheres}},\ }\href
  {https://doi.org/10.1039/c001904f} {\bibfield  {journal} {\bibinfo  {journal}
  {Soft Matter}\ }\textbf {\bibinfo {volume} {6}},\ \bibinfo {pages} {2975}
  (\bibinfo {year} {2010})}\BibitemShut {NoStop}%
\bibitem [{\citenamefont {O'Hern}\ \emph {et~al.}(2002)\citenamefont {O'Hern},
  \citenamefont {Langer}, \citenamefont {Liu},\ and\ \citenamefont
  {Nagel}}]{OHern2002}%
  \BibitemOpen
  \bibfield  {author} {\bibinfo {author} {\bibfnamefont {C.~S.}\ \bibnamefont
  {O'Hern}}, \bibinfo {author} {\bibfnamefont {S.~A.}\ \bibnamefont {Langer}},
  \bibinfo {author} {\bibfnamefont {A.~J.}\ \bibnamefont {Liu}},\ and\ \bibinfo
  {author} {\bibfnamefont {S.~R.}\ \bibnamefont {Nagel}},\ }\bibfield  {title}
  {\bibinfo {title} {{Random Packings of Frictionless Particles}},\ }\href
  {https://doi.org/10.1103/PhysRevLett.88.075507} {\bibfield  {journal}
  {\bibinfo  {journal} {Phys. Rev. Lett.}\ }\textbf {\bibinfo {volume} {88}},\
  \bibinfo {pages} {075507} (\bibinfo {year} {2002})}\BibitemShut {NoStop}%
\bibitem [{\citenamefont {Solon}\ \emph {et~al.}(2015)\citenamefont {Solon},
  \citenamefont {Fily}, \citenamefont {Baskaran}, \citenamefont {Cates},
  \citenamefont {Kafri}, \citenamefont {Kardar},\ and\ \citenamefont
  {Tailleur}}]{Solon2015a}%
  \BibitemOpen
  \bibfield  {author} {\bibinfo {author} {\bibfnamefont {A.~P.}\ \bibnamefont
  {Solon}}, \bibinfo {author} {\bibfnamefont {Y.}~\bibnamefont {Fily}},
  \bibinfo {author} {\bibfnamefont {A.}~\bibnamefont {Baskaran}}, \bibinfo
  {author} {\bibfnamefont {M.~E.}\ \bibnamefont {Cates}}, \bibinfo {author}
  {\bibfnamefont {Y.}~\bibnamefont {Kafri}}, \bibinfo {author} {\bibfnamefont
  {M.}~\bibnamefont {Kardar}},\ and\ \bibinfo {author} {\bibfnamefont
  {J.}~\bibnamefont {Tailleur}},\ }\bibfield  {title} {\bibinfo {title}
  {{Pressure is not a state function for generic active fluids}},\ }\href
  {https://doi.org/10.1038/nphys3377} {\bibfield  {journal} {\bibinfo
  {journal} {Nature Physics}\ }\textbf {\bibinfo {volume} {11}},\ \bibinfo
  {pages} {673} (\bibinfo {year} {2015})}\BibitemShut {NoStop}%
\bibitem [{\citenamefont {Speck}\ \emph {et~al.}(2014)\citenamefont {Speck},
  \citenamefont {Bialk{\'e}}, \citenamefont {Menzel},\ and\ \citenamefont
  {L{\"o}wen}}]{speck2014effective}%
  \BibitemOpen
  \bibfield  {author} {\bibinfo {author} {\bibfnamefont {T.}~\bibnamefont
  {Speck}}, \bibinfo {author} {\bibfnamefont {J.}~\bibnamefont {Bialk{\'e}}},
  \bibinfo {author} {\bibfnamefont {A.~M.}\ \bibnamefont {Menzel}},\ and\
  \bibinfo {author} {\bibfnamefont {H.}~\bibnamefont {L{\"o}wen}},\ }\bibfield
  {title} {\bibinfo {title} {Effective cahn-hilliard equation for the phase
  separation of active brownian particles},\ }\href
  {https://doi.org/https://doi.org/10.1103/PhysRevLett.112.218304} {\bibfield
  {journal} {\bibinfo  {journal} {Physical Review Letters}\ }\textbf {\bibinfo
  {volume} {112}},\ \bibinfo {pages} {218304} (\bibinfo {year}
  {2014})}\BibitemShut {NoStop}%
\bibitem [{SM()}]{SM}%
  \BibitemOpen
  \href@noop {} {}\bibinfo {note} {See Supplemental Material at http://... for
  additional information.}\BibitemShut {Stop}%
\bibitem [{\citenamefont {Wysocki}\ \emph {et~al.}(2014)\citenamefont
  {Wysocki}, \citenamefont {Winkler},\ and\ \citenamefont
  {Gompper}}]{Wysocki2014}%
  \BibitemOpen
  \bibfield  {author} {\bibinfo {author} {\bibfnamefont {A.}~\bibnamefont
  {Wysocki}}, \bibinfo {author} {\bibfnamefont {R.~G.}\ \bibnamefont
  {Winkler}},\ and\ \bibinfo {author} {\bibfnamefont {G.}~\bibnamefont
  {Gompper}},\ }\bibfield  {title} {\bibinfo {title} {{Cooperative motion of
  active Brownian spheres in three-dimensional dense suspensions}},\ }\href
  {https://doi.org/10.1209/0295-5075/105/48004} {\bibfield  {journal} {\bibinfo
   {journal} {EPL (Europhysics Letters)}\ }\textbf {\bibinfo {volume} {105}},\
  \bibinfo {pages} {48004} (\bibinfo {year} {2014})}\BibitemShut {NoStop}%
\bibitem [{\citenamefont {Bialk{\'{e}}}\ \emph {et~al.}(2015)\citenamefont
  {Bialk{\'{e}}}, \citenamefont {Siebert}, \citenamefont {L{\"{o}}wen},\ and\
  \citenamefont {Speck}}]{Bialke2015}%
  \BibitemOpen
  \bibfield  {author} {\bibinfo {author} {\bibfnamefont {J.}~\bibnamefont
  {Bialk{\'{e}}}}, \bibinfo {author} {\bibfnamefont {J.~T.}\ \bibnamefont
  {Siebert}}, \bibinfo {author} {\bibfnamefont {H.}~\bibnamefont
  {L{\"{o}}wen}},\ and\ \bibinfo {author} {\bibfnamefont {T.}~\bibnamefont
  {Speck}},\ }\bibfield  {title} {\bibinfo {title} {{Negative Interfacial
  Tension in Phase-Separated Active Brownian Particles}},\ }\href
  {https://doi.org/10.1103/PHYSREVLETT.115.098301/FIGURES/4/MEDIUM} {\bibfield
  {journal} {\bibinfo  {journal} {Physical Review Letters}\ }\textbf {\bibinfo
  {volume} {115}},\ \bibinfo {pages} {098301} (\bibinfo {year}
  {2015})}\BibitemShut {NoStop}%
\bibitem [{\citenamefont {Berthier}(2014)}]{Berthier2014}%
  \BibitemOpen
  \bibfield  {author} {\bibinfo {author} {\bibfnamefont {L.}~\bibnamefont
  {Berthier}},\ }\bibfield  {title} {\bibinfo {title} {{Nonequilibrium Glassy
  Dynamics of Self-Propelled Hard Disks}},\ }\href
  {https://doi.org/10.1103/PhysRevLett.112.220602} {\bibfield  {journal}
  {\bibinfo  {journal} {Physical Review Letters}\ }\textbf {\bibinfo {volume}
  {112}},\ \bibinfo {pages} {220602} (\bibinfo {year} {2014})}\BibitemShut
  {NoStop}%
\bibitem [{\citenamefont {Morse}\ \emph {et~al.}(2021)\citenamefont {Morse},
  \citenamefont {Roy}, \citenamefont {Agoritsas}, \citenamefont {Stanifer},
  \citenamefont {Corwin},\ and\ \citenamefont {Manning}}]{Morse2021}%
  \BibitemOpen
  \bibfield  {author} {\bibinfo {author} {\bibfnamefont {P.~K.}\ \bibnamefont
  {Morse}}, \bibinfo {author} {\bibfnamefont {S.}~\bibnamefont {Roy}}, \bibinfo
  {author} {\bibfnamefont {E.}~\bibnamefont {Agoritsas}}, \bibinfo {author}
  {\bibfnamefont {E.}~\bibnamefont {Stanifer}}, \bibinfo {author}
  {\bibfnamefont {E.~I.}\ \bibnamefont {Corwin}},\ and\ \bibinfo {author}
  {\bibfnamefont {M.~L.}\ \bibnamefont {Manning}},\ }\bibfield  {title}
  {\bibinfo {title} {{A direct link between active matter and sheared granular
  systems}},\ }\href {https://doi.org/10.1073/PNAS.2019909118/-/DCSUPPLEMENTAL}
  {\bibfield  {journal} {\bibinfo  {journal} {Proceedings of the National
  Academy of Sciences of the United States of America}\ }\textbf {\bibinfo
  {volume} {118}},\ \bibinfo {pages} {e2019909118} (\bibinfo {year}
  {2021})}\BibitemShut {NoStop}%
\bibitem [{\citenamefont {Ciamarra}\ and\ \citenamefont
  {Coniglio}(2009)}]{ciamarra2009jamming}%
  \BibitemOpen
  \bibfield  {author} {\bibinfo {author} {\bibfnamefont {M.~P.}\ \bibnamefont
  {Ciamarra}}\ and\ \bibinfo {author} {\bibfnamefont {A.}~\bibnamefont
  {Coniglio}},\ }\bibfield  {title} {\bibinfo {title} {Jamming at zero
  temperature, zero friction, and finite applied shear stress},\ }\href
  {https://doi.org/https://doi.org/10.1103/PhysRevLett.103.235701} {\bibfield
  {journal} {\bibinfo  {journal} {Physical Review Letters}\ }\textbf {\bibinfo
  {volume} {103}},\ \bibinfo {pages} {235701} (\bibinfo {year}
  {2009})}\BibitemShut {NoStop}%
\bibitem [{\citenamefont {Alert}\ \emph {et~al.}(2022)\citenamefont {Alert},
  \citenamefont {Casademunt},\ and\ \citenamefont {Joanny}}]{Alert2022}%
  \BibitemOpen
  \bibfield  {author} {\bibinfo {author} {\bibfnamefont {R.}~\bibnamefont
  {Alert}}, \bibinfo {author} {\bibfnamefont {J.}~\bibnamefont {Casademunt}},\
  and\ \bibinfo {author} {\bibfnamefont {J.-F.}\ \bibnamefont {Joanny}},\
  }\bibfield  {title} {\bibinfo {title} {{Active Turbulence}},\ }\bibfield
  {journal} {\bibinfo  {journal} {Annual Review of Condensed Matter Physics}\
  }\textbf {\bibinfo {volume} {13}},\ \href
  {https://doi.org/10.1146/annurev-conmatphys-082321-035957}
  {10.1146/annurev-conmatphys-082321-035957} (\bibinfo {year}
  {2022})\BibitemShut {NoStop}%
\bibitem [{\citenamefont {Zheng}\ \emph {et~al.}(2021)\citenamefont {Zheng},
  \citenamefont {Parmar},\ and\ \citenamefont {Ciamarra}}]{Zheng2021}%
  \BibitemOpen
  \bibfield  {author} {\bibinfo {author} {\bibfnamefont {Y.}~\bibnamefont
  {Zheng}}, \bibinfo {author} {\bibfnamefont {A.~D.~S.}\ \bibnamefont
  {Parmar}},\ and\ \bibinfo {author} {\bibfnamefont {M.~P.}\ \bibnamefont
  {Ciamarra}},\ }\bibfield  {title} {\bibinfo {title} {{Hidden Order Beyond
  Hyperuniformity in Critical Absorbing States}},\ }\href
  {https://doi.org/10.1103/PhysRevLett.126.118003} {\bibfield  {journal}
  {\bibinfo  {journal} {Physical Review Letters}\ }\textbf {\bibinfo {volume}
  {126}},\ \bibinfo {pages} {118003} (\bibinfo {year} {2021})}\BibitemShut
  {NoStop}%
\end{thebibliography}

%


\newpage
\setcounter{figure}{0}
\setcounter{equation}{0}
\newcommand{\sFrac}[2]{{\textstyle\frac{#1}{#2}}}
\def\u0#1{\underline {#1}}
\def\theequation{S\arabic{equation}}
\renewcommand*{\thefigure}{S\arabic{figure}}

\onecolumngrid

\section{Motility Induced Phase separation}
\begin{figure}[!h]
\centering
\includegraphics[angle=0,width=0.42\textwidth]{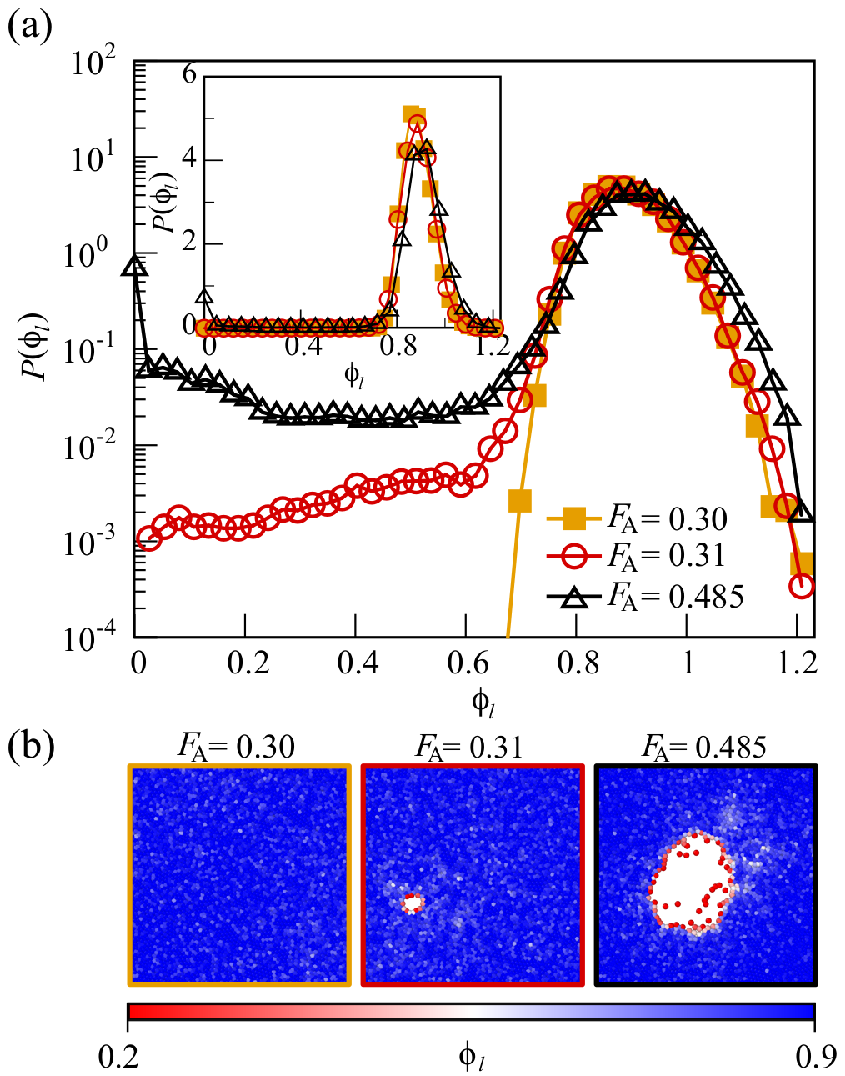}
\caption{
(a) Probability distribution of the local volume fraction at $\phi = 0.89$ and different $F_A$ values. We consider the system as phase separated when this distribution is bimodal, e.g., $F = 0.485$ or when it possesses an extended left tails, e.g., $F = 0.31$. Conversely, the system in in the homogeneous phase, e.g., $F = 0.30$.
(b) illustration of a fraction of the systems with
particles color coded according to their local volume fraction, $a_i/s_i$, where $a_i$ is the particle area, and $s_i$ the area of its Voronoi cell evaluated via a radical tessellation of the system. 
The peak in zero of $P(\phi)$ (panel a) and the system's illustrations (panel b) clarify that the coexisting phases are a dense liquid and an empty gas with $\phi = 0$.
\label{fig:probphi}
}
\end{figure}

We consider the system to be  phase-separated when $P(\phi_l)$ has a bimodal distribution or has a long tail extending to low densities.
Indeed, the direct visualization of the system clarifies that long tails occur in the presence of tiny `gas bubbles' within the system, as in Fig.~\ref{fig:probphi}.
We consider these bubbles a signal of phase separation as we expect them to merge and make the area fraction distribution bimodal via a coarsening process.
At the considered high-density values, this coarsening dynamics is too slow to be numerically followed.

\section{From MIPS to unjamming/jamming}
\begin{figure}[!h]
\centering
\includegraphics[angle=0,width=0.7\textwidth]{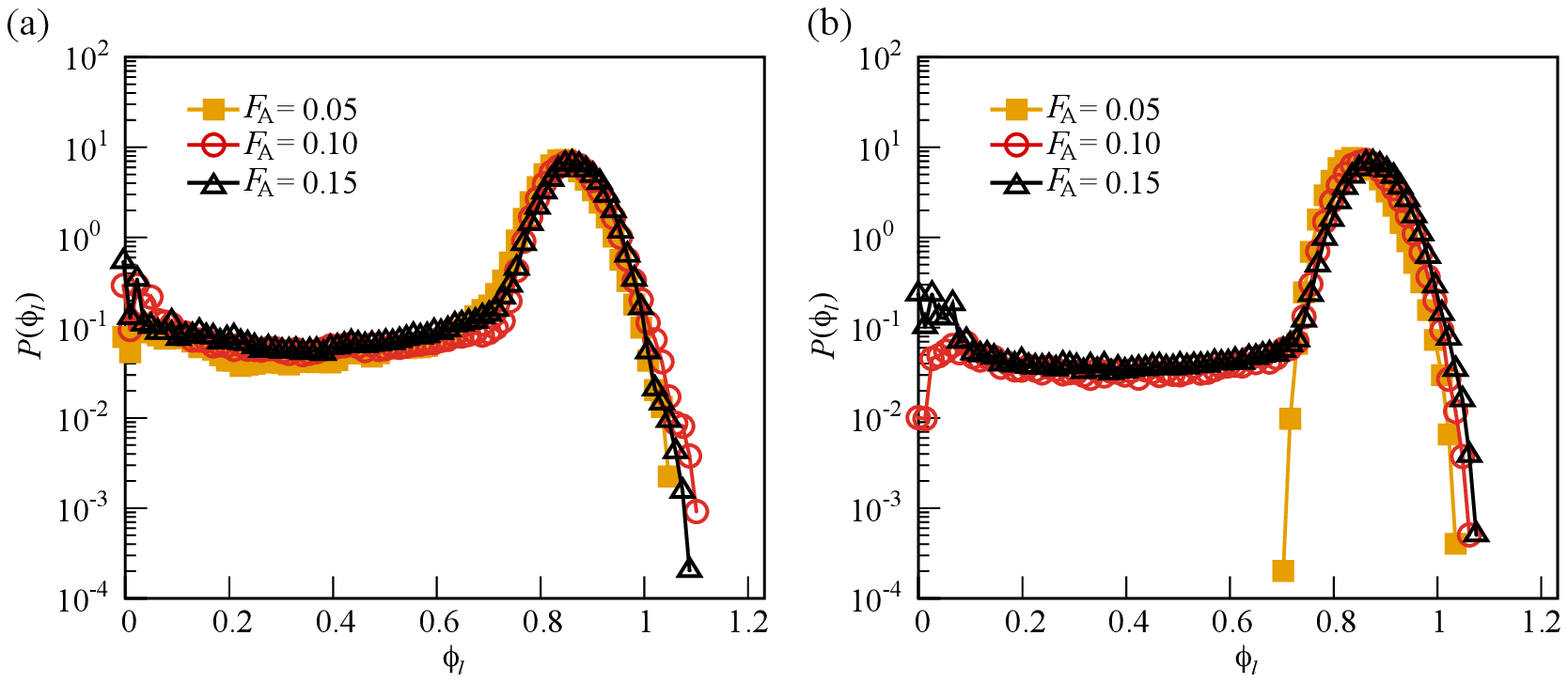}
\caption{
Active force dependence of the probability distribution of the local volume fraction at (a) $\phi = 0.82 < \phi_J$ and (b) $\phi = 0.845 > \phi_J$.
\label{fig:probphi_jamming}
}
\end{figure}

We clarify how a phase-separated system becomes unjammed or jammed as the magnitude of the active force decreases by illustrating the force dependence of the distribution of the local volume fraction for densities below and above the jamming transition, respectively in Fig.~\ref{fig:probphi_jamming}(a) and (b).

Regardless of the density value, at high values of the active force in the MIPS region, the density distribution is bimodal and in the gas phase $\phi \simeq 0$. 
At the considered densities, the system appears as a liquid with cavities, e.g., like the one illustrated in Fig.\ref{fig:probphi}(b) for $F_{\rm A}=0.485$.

For $\phi < \phi_J$, the system reaches the hard-sphere limit as the magnitude of the active force decreases. 
Consistently, the density distributions of Fig.~\ref{fig:probphi_jamming}(a) become active-force independent, and in the $F_A \to 0$ limit, the system converges to an unjammed inhomogeneous state.

For $\phi > \phi_J$, cavities shrink as the active force decreases.
Indeed, we see a drop in the $\phi \simeq 0$ peak of the density distribution as $F_A$ varies from $0.15$ to $0.10$.
Cavities disappear for small enough forces, and the system becomes jammed and homogeneous ($F = 0.05$).

\section{Dynamics within the coexistence region}
\begin{figure}[!h]
\centering
\includegraphics[angle=0,width=0.42\textwidth]{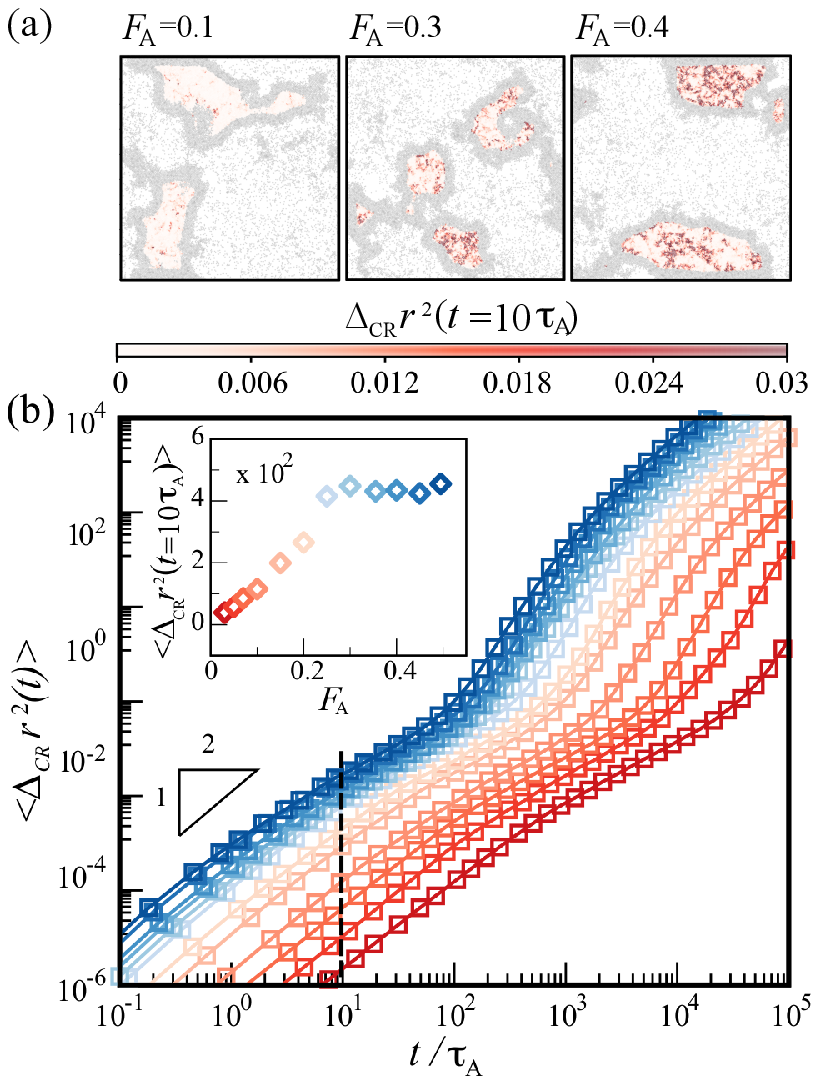}
\caption{
We identify particles of the dense phase in the coexistence region via thresholding criterion on the local density.
Panel (a) shows maps of their cage-relative mean square displacement at time $t = 10\tau_A$ and different values of $F_A$. Gas-like particles are in grey.
Panel (b) illustrates the cage-relative mean square displacement of the dense phase for different $F_A$ values as a function of $t/\tau_A$.
The inset shows the $F_A$ dependence of the mean-square displacement at $t = 10\tau_A$.
\label{fig:ps_dynamics}
}
\end{figure}

Within the coexistence region the system phase separates in a dilute and a dense phase.
In monodisperse systems of active persistent particle, previous results showed the dense phase to be crystalline (C. Reichhardt and C.J. O. Reichhardt, Soft Matter {\bf 19}, 7502, 2014).
We have investigated the dynamics of the dense phase and found it to be of liquid type. 
To this end, we identify the particles of the dense phase via a threshold criterion on a particle-defined local density, at time $t = 0$.
An example of this approach is in Fig.~\ref{fig:ps_dynamics}(a), where
particles are color coded if belonging to dense phase, and grey if belonging to the gas phase.

We have then investigated the cage-relative mean square displacement of the particles of the dense phase.
Fig.~\ref{fig:ps_dynamics}(b) illustrates results at $\phi = 0.41$, for different values of the active force (see inset) in the coexistence region.
Time is in unit of the characteristic time scale fixed by the active velocity, $\tau_A$.
The mean square displacement exhibits a crossover towards an asymptotic ballistic behavior, which is apparent at large enough $F_A$.
This crossover signals the disruption of the dense cluster.
The behavior of the mean square displacement proves that the dense phase is of liquid type.

Fig.~\ref{fig:ps_dynamics}(b) further shows that the active force does not simply set the time scale of the system.
Indeed, if $F_A$ were only affecting the typical time scale $\tau_{A}$, then data for different $F_A$ values would collapse when plotted versus $t/\tau_{AA}$.
In Fig.~\ref{fig:ps_dynamics}(b), we see that this occurs at large enough $F_A$.
Conversely, as $F_A$ decreases, the slowing down of the dynamics is faster than that of $1/\tau_A$.
This result is apparent from the $F_A$ dependence of the mean square displacement evaluated at $t = 10\tau_A$ illustrated in the inset of (b).
The direct visualization of the magnitude of the cage-relative square displacements at $t = 10\tau_A$ and different values of the active force in Fig.~\ref{fig:ps_dynamics}(a) confirms this finding.

\end{document}